\documentclass[fleqn,usenatbib]{mnras}

\usepackage{newtxtext,newtxmath}

\usepackage[T1]{fontenc}
\usepackage{ae,aecompl}

\usepackage{graphicx}	% Including figure files
\usepackage{amsmath}	% Advanced maths commands
\usepackage{amssymb}	% Extra maths symbols
\usepackage{multirow} % using multirow for confusion matrices
\usepackage{pdflscape}
\usepackage{epsfig}

\newcommand{\Chandra}{${\it Chandra}$}
\newcommand{\Hubble}{${\it Hubble}$}

\newcommand{\nustar}{${\it NuSTAR}$}

\newcommand{\changed}[1]{#1}

\title[Random Forest Classification of M31 X-ray Sources]{Identifying New X-ray Binary Candidates in M31 using Random Forest Classification}

\author[R. M. Arnason et al.]{
R. M. Arnason$^{1}$\thanks{E-mail: rarnaso@uwo.ca},
P. Barmby$^{1,2}$,
N. Vulic$^{1,3,4}$
\\
$^{1}$Department of Physics and Astronomy, University of Western Ontario, 1151 Richmond Street, London, ON N6A 3K7, Canada \\
$^{2}$Institute for Earth and Space Exploration, University of Western Ontario, 1151 Richmond Street, London, ON N6A 3K7, Canada \\
$^{3}$Laboratory for X-ray Astrophysics, Code 662, NASA Goddard Space Flight Center, Greenbelt, MD 20771, USA \\
$^{4}$Department of Astronomy and Center for Space Science and Technology (CRESST), University of Maryland, College Park, MD 20742-2421, USA \\
}

\date{Accepted XXX. Received YYY; in original form ZZZ}

\pubyear{2020}

\begin{document}
\label{firstpage}
\pagerange{\pageref{firstpage}--\pageref{lastpage}}
\maketitle

% Abstract of the paper
\begin{abstract}
Identifying X-ray binary (XRB) candidates in nearby galaxies requires distinguishing them from possible contaminants including foreground stars and background active galactic nuclei. 
This work investigates the use of supervised machine learning algorithms to identify high-probability X-ray binary candidates.
Using a catalogue of 943 \Chandra\ X-ray sources in the Andromeda galaxy, we trained and tested several classification algorithms using the X-ray properties of 163 sources with previously known types. 
Amongst the algorithms tested, we find that random forest classifiers give the best performance and work better in a binary classification (XRB/non-XRB) context compared to the use of multiple classes. 
Evaluating our method by comparing with classifications from visible-light and hard X-ray observations as part of the Panchromatic Hubble Andromeda Treasury, we find compatibility at the 90\% level, although we caution that the number of source in common is rather small.
The estimated probability that an object is an X-ray binary agrees well between the random forest binary and multiclass approaches
and we find that the classifications with the highest confidence are in the X-ray binary class. The most discriminating X-ray bands for classification are the 1.7--2.8, 0.5--1.0, 2.0--4.0, and 2.0--7.0~keV photon flux ratios. 
Of the 780 unclassified sources in the Andromeda catalogue, we identify 16 new high-probability X-ray binary candidates and tabulate their properties for follow-up.
\end{abstract}


\section{Introduction}

\changed{
The non-nuclear X-ray emission of galaxies is dominated by X-ray binaries, relatively rare systems that contain a compact object in a close binary orbit with an ordinary star.
The majority of X-ray binaries are classified based on the type of the companion star. 
Compact objects which accrete from a $<1$~M$_{\sun}$ companion undergoing Roche Lobe overflow are known as low-mass X-ray binaries (LMXBs), while compact objects accreting from a $\geq10$~M$_{\sun}$ star, usually through the stellar wind, are identified as high-mass X-ray binaries (HMXBs) \citep{Casares17a}. 
X-ray binaries can also be categorized by the type of compact object accreting material from the companion, either a black hole (BH) or neutron star (NS). 
}

\changed{
Aside from their value as laboratories for extreme physics, XRBs can be used as tracers of galaxy properties.
The X-ray luminosity functions (XLFs) of sources within nearby star-forming galaxies are dominated by contributions from HMXBs.
When normalized by the parent galaxy's star formation rate,  XLFs occupy a narrow band in $N(>L)-L$ space \citep{Grimm03a}.
This trend appears consistent at resolved scales: in the Milky Way, HMXBs cluster spatially close to known active star-forming complexes in the Milky Way's spiral arms \citep{Bodaghee12a}.
Instead of current star formation, LMXBs trace past star formation (via a galaxy's stellar mass) and current stellar density. 
Many LMXBs are found in the globular clusters of galaxies where they can be created through dynamical encounters enabled by the high stellar densities \citep{Verbunt06a}.
Consequently, the fraction of a galaxy's LMXBs that are found in globular clusters increases with the specific frequency of globular clusters \citep{Maccarone03a}. 
In addition, low mass stars comprise the bulk of the stellar mass in any stellar population, so LMXBs within a galaxy can trace their host galaxy's stellar mass \citep{Gilfanov04a}.
}

\subsection{Identifying X-Ray Binaries}

\changed{
While understanding the physics of X-ray binaries is best accomplished through detailed study of Milky Way sources, population studies of X-ray sources in the Milky Way are challenging.
Distances to individual sources can be highly uncertain \citep{Gandhi2019} and dust and gas in the disk obscures our view along important lines of sight.
External galaxies have all sources at effectively the same distance.  
We can resolve the structure of nearby galaxies at a favourable viewing angle while still detecting a substantial fraction of their X-ray source populations.
The advent of high spatial resolution, sensitive X-ray observations in the \textit{Chandra} era has permitted the study of X-ray sources in nearby galaxies on resolved scales. 
}

\changed{
To use XRBs as a probe for galaxy structure and properties, accurate determinations of the XRB count in a population are required. 
Nearby galaxies, with their large angular sizes, can have X-ray source lists contaminated by X-ray active foreground stars in the Milky Way or background AGN. 
Depending on the X-ray energy bands used, supernova remnants in the same galaxy can mimic the appearance of X-ray binaries \citep{Grimm03a}. 
The preferred method to resolve this source confusion is to identify infrared or visible-light counterparts, where spectroscopy can separate AGN from expected XRB counterparts. 
In addition, sources that are spatially extended at longer wavelengths can be identified as associated with distant galaxies and therefore likely AGN, while sources that have high proper motion or an extreme optical to X-ray flux ratio can be identified as foreground stars \citep{Vilhu87a,Guillot09a,Saeedi16a}.
Multiwavelength observations of X-ray binary candidates may not be available or practical due to crowding, extinction, or large distance. 
}

\begin{figure}
\centering
\includegraphics[angle=-90, width=1.0\columnwidth]{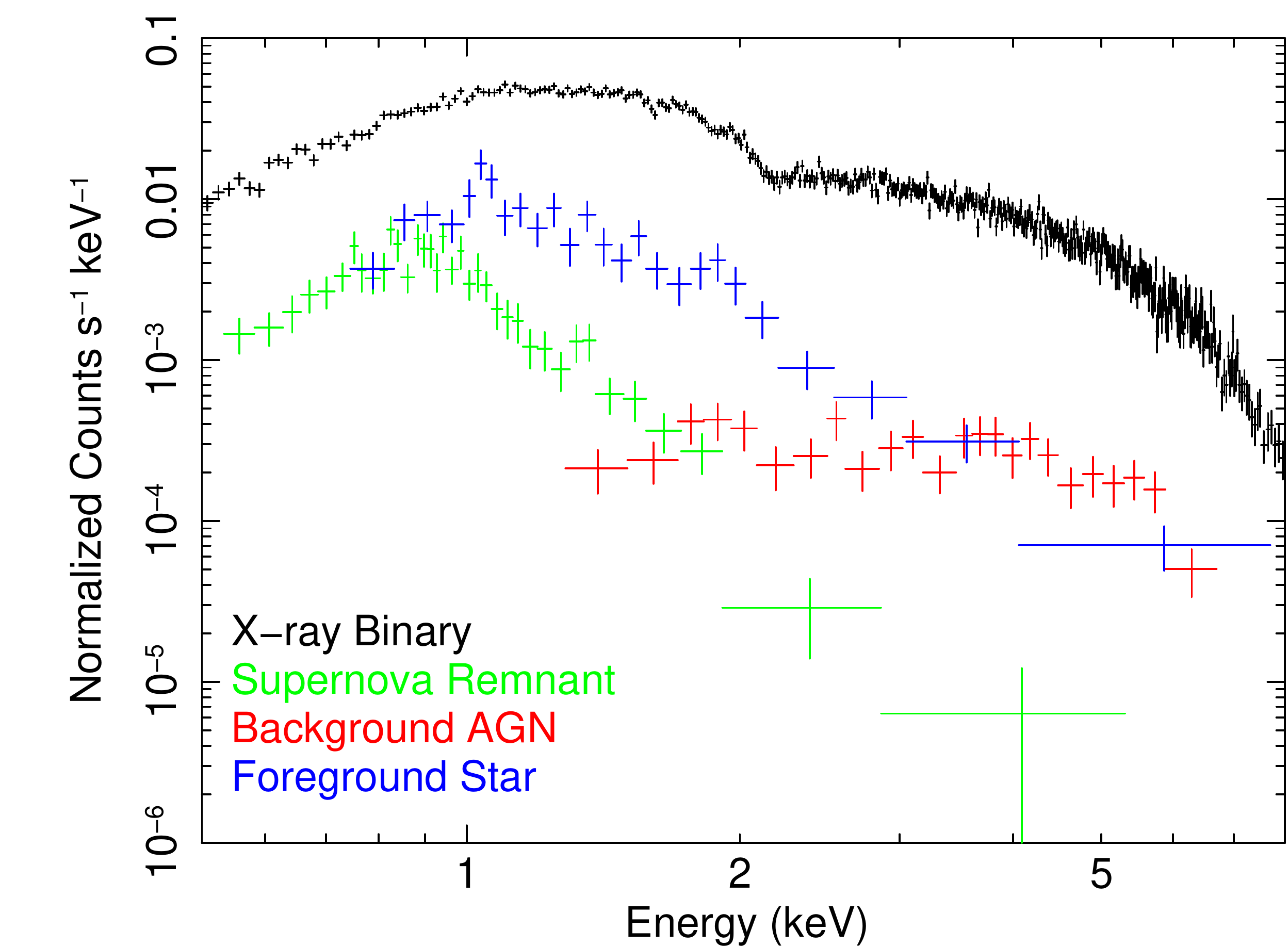} 
\caption{\Chandra~X-ray spectra of point source types detected in the direction of M31. The spectral shape of each source type is unique across the \Chandra\ energy band of $0.5-8.0$ keV, assuming sufficient source counts. 
In the low-count regime, advanced techniques such as ML are required to differentiate sources.}
\label{fig:xrspec}
\end{figure}

\changed{
In the absence of multiwavelength observations, differentiating bright extragalactic XRBs from background AGN, foreground stars (fgStars), and supernova remnants (SNRs) can be accomplished using the unique signatures in their X-ray spectra. 
In \autoref{fig:xrspec} we show an X-ray spectrum for the most luminous source of each type in the direction of M31, where data have been taken from \citet{Vulic16a} and \citet{Williams18a}.
XRBs detected in nearby galaxies are generally well-described by an absorbed power-law with spectral index $\Gamma\approx1.7$ \citep[e.g.,][]{colbert02-04}, although this depends on numerous physical properties that affect their observed spectral states \citep[for a detailed review, see][]{done12-07}. 
 XRBs in different spectral states have been successfully identified in nearby galaxies with the inclusion of \nustar\ data; 
the additional spectral curvature in the hard energy band at $10<E<25$~keV can serve to identify XRB types and differentiate them from AGN \citep{Wik14a,Yukita16a,Vulic18a,lazzarini2018}.
This type of classification is beyond the scope of this paper and is only feasible in the nearest galaxies because of the lower angular resolution of hard X-ray observations.
}

\changed{
AGN residing in resolved galaxies can readily identified in high-resolution optical images that show the galaxy as an extended source.
With only X-ray data, simple techniques may or may not distinguish AGN and XRBs.
AGN are known to have power-law spectral indices similar to XRBs, with $\Gamma\approx1.7$ for unobscured AGN \citep[e.g.,][]{svoboda07-17}, whereas heavily absorbed AGN have lower values of $\Gamma$. 
The background AGN spectrum shown in \autoref{fig:xrspec} has a power-law slope of 1.4, and is thus easily distinguished from the XRB spectrum. 
For unabsorbed AGN, or low-count sources in which spectral fitting is not possible, different approaches are necessary.
}

\changed{
SNRs and fgStars both have soft X-ray spectra that separates them from XRBs and AGN. 
SNRs are separated into two classes: sources that have a thermal component dominating the X-ray spectrum at $E<2$ keV (shell-like) and power-law dominated sources (Crab-like; pulsar wind nebulae) \citep{vink12-12}. 
SNRs are typically very soft sources, such as the example shown in  \autoref{fig:xrspec}, which is described by a shocked ISM component with temperature $kT=0.24$ keV. 
X-ray emission has been detected from both early-type and late-type stars \citep{schmitt-00, gudel09-09}, however, X-ray emission from fgStars is usually attributed to flares from late-type stars \citep[e.g., M-dwarfs;][]{guo02-16,tsang08-12}. 
The X-ray spectra of flare stars are best fit with one-temperature or two-temperature plasma emission models. 
The fgStar in \autoref{fig:xrspec} was best fit with a two-temperature model having $kT_{1}=0.6$ keV and $kT_{2}=2.4$ keV.
}

\changed{
If neither  multiwavelength observations or X-ray spectra are available, classification of extragalactic X-ray sources 
can make use of the presence of X-ray features unique to compact objects (or neutron stars in particular), such as Type I X-ray bursts or X-ray pulsations.
If time-resolved data are not available, then classification is typically done using a combination of X-ray brightness and X-ray colours or hardness ratio. 
X-ray colour-colour diagrams or colour-colour-intensity diagrams in the $0.5-10$~keV bands show that certain types of objects tend to cluster together based on, for example, compact object type and pulsating versus non-pulsating neutron stars \citep{Prestwich03a,Vrtilek13a}; 
as discussed above, adding hard X-ray data can help separate some X-ray source types, for example black hole versus neutron star binaries \citep{Vulic18a}. 
Sensitivity limits mean that only the brightest sources are detected in hard X-rays, so the utility of hard X-ray data for classifying large samples of extragalactic X-ray sources is limited.
Colour-colour diagrams or colour-colour-intensity diagrams made from observations at softer X-ray energies, such as with \textit{Chandra}, usually result in approximate decision boundaries that are difficult to constrain.
}

\subsection{Machine Learning for X-ray Source Classification} 
One approach to overcoming the approximate decision boundaries found through simple colour-colour or hardness-intensity diagrams is to 
apply machine learning techniques to make optimal use of the information available in low-energy-resolution X-ray data.
\changed{Machine learning (ML) can be either supervised or unsupervised; see \citep{Baron19} for an overview of applications in astronomy.
In supervised machine learning a function/algorithm ``learns" outputs based on a previously trained set of paired input/output data.}
These techniques can predict either as a regression or a classification when applied to new, unknown examples.
Machine learning classifiers tend to perform optimally on large datasets with many classified examples and a sufficient number of informative features which can help define a model that discriminates between different desired categories. 

In astronomy, machine learning has already been used to investigate classification problems where the data may have high dimensionality and is difficult to either model or assign simple classification boundaries.
Recently, \cite{Ksoll18a} used machine learning techniques to separate lower main sequence stars from pre-main sequence stars using a random forest algorithm applied to the \textit{Hubble Space Telescope} (\textit{HST}) photometry of 30 Doradus. 
Machine learning has previously been used to investigate the properties of X-ray binaries.
\cite{Sonbas16} used a learning decision tree algorithm to classify X-ray sources in the Draco dwarf galaxy
\changed{on the basis of X-ray fluxes in four bands and visible through mid-infrared counterpart photometry.
They found that} classifications made by their method tend to agree with classifications made with previously established classification techniques \changed{\citep[e.g., by][]{Saeedi16a}}.
\cite{Gopalan15a} expanded the colour-colour-intensity diagram classification technique by applying a supervised learning algorithm as a method of demarcating systems containing black holes, pulsating neutron stars, or non-pulsating neutron stars. 
\changed{\citet{Lo14} employed the random forest algorithm to classify time-varying X-ray sources in the Second {\em XMM-Newton} Serendipitous Source Catalog using X-ray photometric time series and spectra and multiwavelength information. 
Their results indicated a high classification accuracy and the ability to detect unusual objects in their test sample.
}

Our goal in this work is to improve on previous investigations of the nature of X-ray sources in nearby galaxies by developing a better method of differentiating extragalactic X-ray binaries from other sources using only their X-ray emission.
\changed{There are two broad science goals that using machine learning classification techniques on X-ray sources can achieve. 
The first is to aid in computing the X-ray luminosity function for a galaxy, where avoiding contamination by non-XRBs is important, but incompleteness is less critical if it can be quantified and corrected. 
The second science goal is identifying new XRB candidates for follow-up, where generating a complete sample is important.
Supervised ML is most appropriate here, since this is a classification problem: our goal is to identify objects that belong to a specific class (X-ray binary candidates) of which previously-classified examples exist. 
We consider a large dataset of \textit{Chandra}-detected X-ray sources in the direction of the Andromeda galaxy, M31, and test multiple supervised machine learning algorithms at once.  
The abundance of multiwavelength data available for M31, such as the \textit{HST} Panchromatic Hubble Andromeda Treasury \citep[PHAT;][]{Dalcanton12a}, allows us to evaluate the efficacy of machine learning algorithms at identifying new XRB candidates by independently assessing source classifications.
}

\section{Data and Method}

\changed{
As the the nearest large galaxy to our own Milky Way, and the closest spiral galaxy, M31 is similar enough to the Milky Way for sensible comparison of the X-ray source populations.
It is viewed at a relatively favourable inclination angle without many of the observational complications inherent to observing the Milky Way itself.
M31 has been extensively observed in X-rays with hundreds of sources detected \citep{Stiele11a,Vulic16a, Williams18a}; its
background AGN population has also been extensively investigated compared to other nearby galaxies \citep[e.g.][]{dw2017,Huo2015,Meusinger2010}. 
This means that M31 X-ray sources represent an excellent test case for machine learning approaches because ML algorithms typically perform much better with higher numbers of classified (and total) examples.
}

\begin{figure}
\includegraphics[width=8cm]{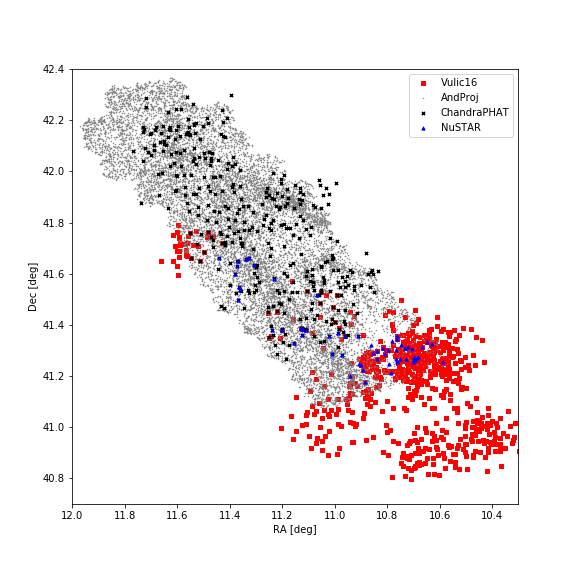}
\caption{
\Chandra\, \Hubble\ and {\it NuSTAR} sources in M31.
Red squares: unclassified \Chandra\ sources from \citet{Vulic16a}, 
grey dots: Andromeda Project \changed{non-stellar} (\textit{HST}) sources from \citet{Johnson15a},
black crosses: \Chandra-PHAT sources from \citet{Williams18a},
blue triangles: {\it NuSTAR}-\Chandra\ sources from \citet{lazzarini2018}.
Not shown here are sources in an additional  \Chandra\ field in the southwest disk \citep[see Fig.~1 of][]{Vulic16a}; these
data were used in our analysis but do not overlap with PHAT.\label{fig:phat_chandra}
}
\end{figure}

\subsection{\Chandra\ Data} 
As our sample dataset, we consider the catalogue of \Chandra\ X-ray sources in M31 from \cite{Vulic16a}.
A full description of the catalogue can be found in that paper, however we summarize the key characteristics here.
This catalogue resulted from combining all 133 available ACIS-I/S observations of M31 to detect sources in three distinct fields (bulge, northeast, and southwest) at a 0.5--8.0 keV luminosity limit of $10^{34}$~erg~s$^{-1}$, covering a total area of $\sim0.6$~deg$^{2}$. 

In total, the catalogue we use  consists of 943 sources.
There are more sources than tabulated in \cite{Vulic16a} because we also include sources that did not meet the ``probability of no-source" cutoff of $1\times10^{-2}$ used as a filter in that study. 
Each source has, in addition to a variety of observational features, the position, median incident energy, mean observed energy, and the photon flux in 16 different energy bands between 0.5--8.0 keV.
These energy bands, the \texttt{ACIS EXTRACT} defaults, can often be represented as linear combinations of other energy bands, and as such they are likely to be highly correlated.
A source only needs to be detected in one of the 16 energy bands to be part of the catalogue, and for most sources at least one of the bands has a flux that is zero or consistent with zero within error.

\begin{table}
\caption{Summary of dataset properties}
\label{tab:dataset}
\begin{tabular}{l|cc}
\hline \hline
Feature Name & \# classified  & \#  unclassified  \\
\hline
0.5 -- 8.0 keV photon flux & 163 & 780 \\
0.5 -- 2.0 keV photon flux fraction & 163 & 749 \\
2.0 -- 8.0 keV photon flux fraction & 153 & 744 \\
0.5 -- 1.7 keV photon flux fraction & 163 & 736 \\
1.7 -- 2.8 keV photon flux fraction & 152 & 679 \\
2.8 -- 8.0 keV photon flux fraction & 147 & 723 \\
0.5 -- 1.5 keV photon flux fraction & 162 & 728 \\
1.5 -- 2.5 keV photon flux fraction & 156 & 684 \\
2.5 -- 8.0 keV photon flux fraction & 149 & 731 \\
0.5 -- 1.0 keV photon flux fraction & 155 & 634 \\
1.0 -- 2.0 keV photon flux fraction & 163 & 719 \\
2.0 -- 4.0 keV photon flux fraction & 148 & 686 \\
4.0 -- 6.0 keV photon flux fraction & 139 & 636 \\
6.0 -- 8.0 keV photon flux fraction & 121 & 513 \\
0.5 -- 7.0 keV photon flux fraction & 163 & 779 \\
2.0 -- 7.0 keV photon flux fraction & 152 & 742 \\
Mean Observed Energy & 163 & 768 \\
Mean Incident Energy & 156 & 664 \\
\hline 
\end{tabular}

\bigskip
The number of classified and unclassified objects per feature varies because some objects have feature values set to zero due to a negative flux or energy being inferred from ACIS EXTRACT.

\end{table}

\subsection{Andromeda Project and Chandra-PHAT Data}
\label{sec:phat_data}
In addition to internal performance metrics, we attempted to validate our technique by comparing classifications from our best-performing algorithms to a classification not based on X-ray source properties.
For this, we used data from the Panchromatic Hubble Andromeda Treasury (PHAT), a large multi-year survey that obtained \textit{HST} photometry for roughly a third of M31's disk in multiple filters across 23 ``bricks" of observations (see \citealt{Dalcanton12a,Williams14a} and accompanying articles). 
This survey permits some 100 million individual stars and other objects of interest to be resolved.
We use the star cluster and background galaxy catalogs of \cite{Johnson15a}, which provide a library of 2,753 clusters and 2,270 background galaxies in the field of M31.
These clusters and background galaxies were catalogued from a citizen-science project that classified objects based on their morphology in PHAT images.

As \autoref{fig:phat_chandra} shows, only a portion of the \citet{Vulic16a} \Chandra\ data has PHAT coverage.
A more comprehensive overlap between \Chandra\ and \Hubble\ M31 observations was provided by the \Chandra-PHAT project \citep{Williams18a}.
This program detected 373 sources to a limiting X-ray flux of  $10^{-15}$~erg~cm$^{-2}$~s$^{-1}$ over an area of $\sim0.4$~deg$^{2}$. 
\citet{Williams18a} found 188 X-ray sources to have counterpart candidates in the \Hubble\ imaging, including 
107 extended background galaxies, 58 point sources potentially associated with M31, 12 foreground stars,
6 supernova remnants and 5 star clusters.
\changed{\citet{Williams18a} suggested that most of the 58 point source counterparts (whose colours are inconsistent with those of foreground stars) are likely to be background galaxies or binaries in M31 and that
X-ray sources without PHAT counterparts are ``most likely to be undetected background galaxies and low-mass X-ray binaries in M31.''}
Companion work by \citet{lazzarini2018} used the \Chandra-PHAT and \Hubble\ observations in conjunction with M31 disk
observations from {\it NuSTAR\/} and an additional \Chandra\ field in the M31 bulge.
These authors identify 15 high-mass X-ray binary candidates and their compact object types, donor star spectral types, and ages.
By design, there is little overlap between the \citet{Vulic16a} and \Chandra-PHAT catalogues, but
there are some sources in common between both these two and the \citet{Vulic16a} and  \citet{lazzarini2018} catalogues.

\subsection{Classification Scheme}
We create a training set by using sources that are classified by \cite{Vulic16a} through crossmatching with sources classified in the literature, principally the \textit{XMM-Newton} source catalogue of \cite{Stiele11a}. 
In total, there are 163 previously classified sources, of which 77 are X-ray binaries (XRB), 43 are background active galactic nuclei (AGN), 29 are fgStars, and 14 are SNRs. 
Since we are primarily interested in the identification of new XRB candidates, we use machine learning algorithms to classify the unknown objects in two ways: first, we consider a multiclass classification where we attempt to classify new objects as XRB, AGN, fgStar, or SNR. 
Secondly, we consider a binary classification where we attempt to determine if an object is an XRB or not.
The multiclass classification allows us to evaluate the viability of classification across multiple object types, while the binary classification allows us to use ML performance metrics (e.g., receiver-operating characteristic curves) that require a two-class formulation of the problem.
Additionally, given the small number of classified sources overall ($<200$), we can expect that a classification scheme with only two categories will perform better.
The binary classification scheme also has a more even distribution of objects between classes, compared to the multiclass scheme.

\subsection{Feature Construction}
In order to evaluate the dataset using machine learning techniques, we must first construct feature inputs for the algorithms.
In order to construct distance-independent features, we divide each of the fluxes by the broad band 0.5--8.0 keV flux.
We therefore use, for each energy band, the ratio of the flux emitted in that particular band to the total flux measured.
Our final list of features consists of the fifteen ratios, the 0.5--8.0 keV flux, and the median incident and observed energy of each source. 
The number of sources which have a non-zero value for each feature in both the classified and unclassified sets is given in \autoref{tab:dataset}. 
As the Table shows, the ratio of classified to unclassified datapoints is about 1:5.

Since uncertainties may result in measured photon fluxes that are negative or that exceed the overall 0.5--8.0 keV photon flux, it is possible that computing flux ratios may yield aphysical values that are negative or greater than 1.
Crowding near the bulge may make 0.5--8.0 keV fluxes unexpectedly smaller because of contamination from nearby sources - \texttt{ACIS EXTRACT} handles crowding by shrinking extraction regions. 
Since \changed{the extraction regions} contain all of the available counts, the 0.5--8.0 keV flux is more likely to be affected by this crowding than other bands.
For physical reasons, we set the photon flux ratios to be zero if they are less than zero, and to one if they are greater than one.

\subsection{Algorithms}
To explore the dataset using machine learning algorithms, we use supervised learning algorithms from the Python \texttt{sklearn} package, version 0.19.1 \citep{sklearnpaper}. 
These algorithms use a set of already-classified training data as input for classifying new data.
Since training and evaluation of a dataset of this size is relatively quick, we evaluated multiple algorithms at once.
Each algorithm used has a number of initialization parameters, also known as ``hyperparameters," which change the fitting behaviour of the algorithm. 
We evaluated each of these hyperparameters by performing a one-dimensional search over gridded values of the hyperparameter to look for the best value. 
The scoring for the best value is specified by the user -- in this case we used the cross-validation score (see \autoref{sec:multi-results} for an explanation) as the determinant of the best estimator.
In many cases, there was no obvious trend for a ``best" value for a given hyperparameter.
In cases where there was a clear ``best" value of the hyperparameter, we use (and specify) that value.
Otherwise, we use the default value of that hyperparameter.
We tested multinomial logistic regression, Gaussian naive Bayes, random forest, a linear support vector classifier, and a multi-layer perceptron neural network.
We chose to evaluate these algorithms since they are commonly used machine learning algorithms for a variety of classification tasks.

Logistic regression is a generalized model-fitting technique similar to linear regression, except that it attempts to fit to probability of class membership instead.
In general, it assumes that class membership is linearly separable in the feature space. 
Based on results from simpler techniques of classifying X-ray sources, such as hardness ratio diagnostics, we do not expect that our categories of X-ray source are linearly separable in the feature space.
However, logistic regression provides a useful baseline comparison and could be considered similar to a simple classification cut made in the feature space.
In our logistic regression model, we used the following \texttt{sklearn} hyperparameters on the \texttt{sklearn.linear\_model.LogisticRegression()} function: balanced class weights, one-versus-rest multi-class handling, L1 penalty with a SAGA (Stochastic Average Gradient with support for L1 regularization) solver, inverse regularization strength $= 1.0$ and a stopping tolerance of 0.001.

Gaussian naive Bayes is a model that produces conditional class probabilities using a Bayesian formulation with the additional assumption that all features are conditionally independent from each other given the class label. 
Since our feature set is not conditionally independent in general, it also provides a useful baseline for comparison.
We trained the naive Bayes algorithm \texttt{sklearn.naive\_bayes.GaussianNB()} using default parameters.

The random forest method uses the aggregate results of an ensemble of decision trees that have been fit on a subset of features and samples.
Each of these decision trees uses a loss function to divide up the samples by segmenting the feature space until all of the ``leaves" contain samples of only one type.
In this case, the loss function is a function that optimizes feature space segmentations (branches of the decision tree) to have samples of only one type with the fewest number of segmentations.
The random forest classifies new samples as the classification returned from the majority of the decision trees in the forest.

Although individual decision trees are highly biased towards the subsample of data/features they fit from, in aggregate the random forest is not strongly biased by its training set.
In addition, random forest algorithms are typically useful even in cases when features are not normalized and when there are a relatively small number of features in the dataset.
We trained the random forest \texttt{sklearn.ensemble.RandomForestClassifier()} using 500 decision trees, with balanced class weights, made splits using an entropy/information gain loss function, and with a maximum tree depth of 80.

Linear support vector classification (SVC) is a technique which fits a separating hyperplane in the feature space that can be used for classification of future examples. 
We fit the \texttt{sklearn} linear SVC \texttt{sklearn.svm.LinearSVC()} using default parameters.

Finally, we used a multi-layer perceptron, which is a class of neural network that learns a non-linear function to classify samples. 
It possesses non-linear hidden layers that learn between the feature inputs and the fitted output.
Multi-layer perceptrons are advantageous in that they learn non-linear functions well, but they often require extensive hyperparameter tuning to be effective.
We used a multi-layer perceptron \texttt{sklearn.neural\_network.MLPClassifier} with 1 hidden layer of 100 neurons, logistic activation, an LBFGS (limited-memory Brodyen-Fletcher-Goldfarb-Schanno algorithm) solver, and a constant learning rate initialized at 0.01.

We also tested the random forest algorithm available through the \textbf{R} randomForest package v. 4.6-12 \citep{randomforestr,Rmanual}.
Given that we find the random forest to have the overall best performance of the sklearn algorithms (see below in Sections~\ref{sec:multi-results} and \ref{sec:binary-results}), we wish to compare the random forest implementations from two of the most popular machine learning packages, expecting that they should give similar performance. 
We used identical hyperparameters to the \texttt{sklearn} random forest (where such hyperparameters could be specified) in order to compare similar realizations of the fitted algorithm.

\subsection{Algorithm evaluation}

We evaluated the algorithms in multiple ways. 
Firstly, we randomly split the classified samples 70\%/30\% into a training and test set, training each algorithm on the majority of the samples and testing them on the remainder.
This training/test split is a relatively common ratio in machine learning problems chosen to avoid overfitting \citep[e.g.,][]{Ksoll18a}.
The accuracy, defined as the number of correct classifications divided by the total number of classifications, was computed on one particular realization of the training/test split.
In addition, we computed the recall \changed{and precision} on this realization of the training/test split, defined per class (e.g., XRB, SNR, AGN, fgStar) as the number of correct classifications of that class divided by the total number of true members of that class \changed{(recall) or the number of correct classifications of that class divided by the total number of classified members of that class (precision). 
For computing the X-ray luminosity function for a galaxy, the precision is our desired metric, since we are looking to avoid contamination but are less concerned with missing XRBs if we are able to estimate the completeness appropriately. 
For the science goal of finding new XRBs in the population, our desired metric is recall, since we would like to avoid missing any candidate XRBs for followup.  }
Since we have significant class imbalance, we computed the recall \changed {and precision} as the average score across each class weighted by the number of true instances of that class.
The accuracy, recall, \changed{and precision} of the trained model as applied to the test set are tabulated in \autoref{tab:r_multi_perf}.
We also computed confusion matrices, which track the predicted versus actual class for all objects in the sample set.

Secondly, we performed $k$-fold cross-validation on the entire classified dataset. 
In $k$-fold cross-validation, the dataset is partitioned into $k$ subsamples.
Each of the $k$ subsamples is used as validation for a model trained on the remaining $k-1$ subsamples, and the cross-validation score is calculated as the mean and standard deviation of the accuracy across each of the $k$ trials. 
In our cross-validation, we chose $k = 5$, as it is typically chosen as an intermediate value between high values (which yield excessively high bias) and low values (which yield excessively high variance). 

\changed{
For the two implementations of the random forest algorithm, we assess the relative importance of different features to the final, trained classification algorithm using the mean decrease in Gini coefficient.
The Gini coefficient measures the homogeneity of a group of objects.
The mean decrease of the Gini coefficient measures the decrease in the homogeneity of objects in each node of the decision tree when a particular feature is removed from all of the trees in the forest \citep{Breiman84a}. 
}

\section{Results}
\subsection{Multiclass Results}
\label{sec:multi-results}

As can be seen from the results in \autoref{tab:r_multi_perf}, performance metrics for the algorithms are generally poor in the multiclass scenario. 
However, based on the metrics (accuracy, recall, \changed{precision,} cross-validation score), the random forest algorithms (both sklearn and R) tend to have the best performance, and the multi-layer perceptron neural network the poorest. 
Naively, if the classes were balanced, we would expect that the accuracy should be $\sim0.25$ for guessing randomly.
However, we have significant class imbalance, with far more XRBs than any other class. 
A randomly guessing algorithm would tend to always pick the most numerous class, which would give an accuracy of $ 77/163\approx0.47$.
Our algorithms predict better than randomly guessing, however this is not a sufficient baseline for evaluating performance, as class imbalance means that models which always predict the majority class (or classes) will perform well. 
The main source of poor performance is not classifications of true X-ray binaries.
In the confusion matrices for the random forest algorithms, shown in Tables~\ref{tab:py_multi_cm} and \ref{tab:r_multi_cm}, \changed{most of the mis-classifications are} of objects other than XRBs,
especially foreground stars and supernova remnants, which are \changed{both less populated than other classes, and more spectrally similar to each other.}
The algorithm also has difficulty classifying AGN, which are classified roughly evenly into the three other categories. 
\changed{The two implementations of the random forest algorithm return slightly different results. 
We can think of two possible explanations for this: one is that there are hyperparameters that cannot be set identically between the two implementations, and the other is that the different result reflects the two algorithms initializing with a different random seed.
As suggested by its name, the results of random forest classification are not entirely deterministic, and this is important for potential users to recognize.
}

\begin{table}
\caption{Algorithm Evaluation, multiclass case \label{tab:r_multi_perf}}
\begin{tabular}{lcccc}
\hline\hline
Algorithm & Accuracy & Precision & Recall & CV Score \\
\hline
Logistic Regression    & 0.55 & 0.53 & 0.55 & $0.54\pm0.04$ \\
Naive Bayes            & 0.57 & 0.54 & 0.57 & $0.52\pm0.07$ \\
Support Vector Class.  & 0.49 & 0.43 & 0.49 & $0.55\pm0.04$ \\
Random Forest (sklearn) & 0.57 & 0.57 & 0.57 & $0.65\pm0.06$ \\
Multi-layer Perceptron NN & 0.57 & 0.60 & 0.57 & $0.52\pm0.08$ \\
Random Forest (R)         & 0.61 & 0.61 & 0.60 & $0.66\pm0.07$ \\
\hline
\end{tabular}
\end{table}

\begin{table}
\caption{Confusion matrix for sklearn random forest, multiclass case \label{tab:py_multi_cm}}
\begin{tabular}{l|l|c|c|c|c|c}
\multicolumn{2}{c}{}&\multicolumn{4}{c}{Actual Class}&\\
\cline{3-6}
\multicolumn{2}{c|}{}&AGN&SNR&fgStar&XRB&\multicolumn{1}{c}{Total}\\
\cline{2-6}
\multirow{4}{*}{Predicted Class}& AGN & $5$ & $0$ & $2$ & $2$ & $9$ \\
\cline{2-6}
& SNR & $0$ & $3$ & $1$ & $0$ & $4$ \\
\cline{2-6}
& fgStar & $3$ & $2$ & $5$ & $0$ & $10$ \\
\cline{2-6}
& XRB & $6$ & $1$ & $4$ & $15$ & $26$ \\
\cline{2-6}
\multicolumn{1}{c}{} & \multicolumn{1}{c}{Total} & \multicolumn{1}{c}{$14$} & \multicolumn{1}{c}{$6$} & \multicolumn{1}{c}{$12$} & \multicolumn{1}{c}{$17$} & \multicolumn{1}{c}{$49$}\\ 
\end{tabular}
\end{table}

\begin{table}
\caption{Confusion matrix for R random forest, multiclass case \label{tab:r_multi_cm}}
\begin{tabular}{l|l|c|c|c|c|c}
\multicolumn{2}{c}{}&\multicolumn{4}{c}{Actual Class}&\\
\cline{3-6}
\multicolumn{2}{c|}{}&AGN&SNR&fgStar&XRB&\multicolumn{1}{c}{Total}\\
\cline{2-6}
\multirow{4}{*}{Predicted Class}& AGN & $6$ & $1$ & $3$ & $2$ & $12$ \\
\cline{2-6}
& SNR & $0$ & $4$ & $1$ & $0$ & $5$ \\
\cline{2-6}
& fgStar & $3$ & $1$ & $6$ & $0$ & $10$ \\
\cline{2-6}
& XRB & $5$ & $0$ & $2$ & $15$ & $22$ \\
\cline{2-6}
\multicolumn{1}{c}{} & \multicolumn{1}{c}{Total} & \multicolumn{1}{c}{$14$} & \multicolumn{1}{c}{$6$} & \multicolumn{1}{c}{$12$} & \multicolumn{1}{c}{$17$} & \multicolumn{1}{c}{$49$}\\ 
\end{tabular}
\end{table}

\subsection{Two-class Results}
\label{sec:binary-results}

\begin{figure}
    \centering
    \includegraphics[width=\columnwidth]{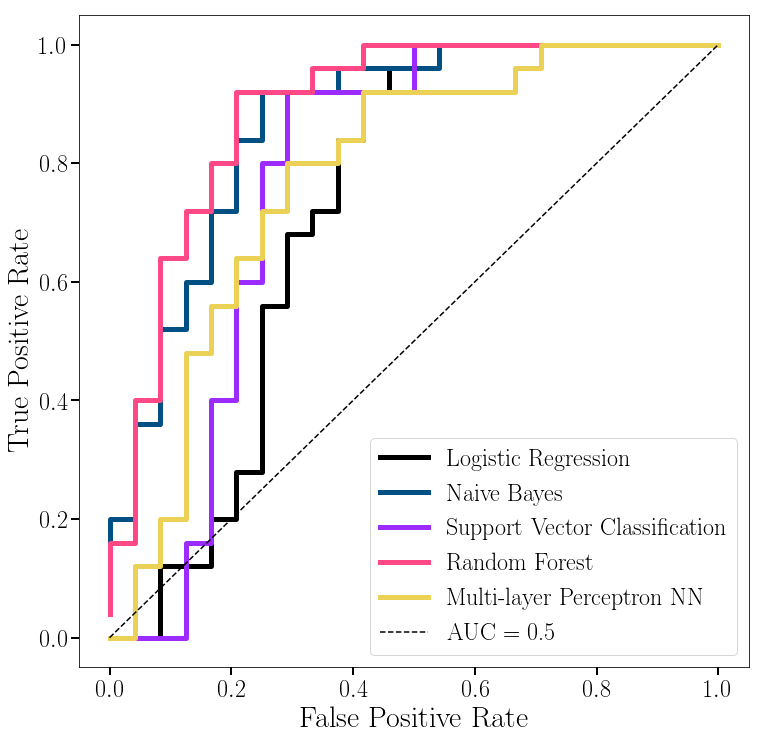}
    \caption{Receiver Operating Characteristic (ROC) curve for all sklearn algorithms when trained on the binary realization of the dataset. An uninformative classifier is plotted for reference.}
    \label{fig:roc_curve}
\end{figure}

We also consider re-evaluation of the problem as a binary one, where we reassign each object to be defined as either an XRB or a non-XRB. 
Since identifying new XRB candidates is the primary goal of this ML-based classification, this permits us to pursue a method that is potentially more accurate overall and is not limited by strong class imbalance.
In addition, using a binary realization of the problem permits us to evaluate algorithm performance using metrics that can only be applied to binary problems, such as the area under the curve (AUC) of a receiver-operating curve.

As with the multiclass case, we performed both a 70/30 training/test split to the classified samples and used the predictions to generate confusion matrices, accuracy, \changed{precision} and recall measures. 
We also performed 5-fold cross-validation on the classified samples and compute the CV score as the mean accuracy across each of the 5 folds.
The scores for each metric are tabulated in \autoref{tab:r_binary_perf}.
In addition, we also create receiver-operating characteristic (ROC) curves for the sklearn algorithms. 
This curve plots the true positive rate against the false positive rate at various classification thresholds for the binary classifier \citep{Spackman89a}.
The area under this curve (AUC) is interpreted as the probability that the classifier will rank a randomly chosen true positive example higher than a randomly chosen true negative example. 
As such, a classifier which guesses randomly (or is uninformative) would have AUC~$= 0.5$.
We plot ROC curves for the sklearn algorithms in \autoref{fig:roc_curve}, and tabulate the AUCs in \autoref{tab:r_binary_perf}.
Each of the algorithms gives an AUC significantly better than an uninformative classifier, though once again the random forest implementations (both sklearn and R) tend to give the best overall performance.

The confusion matrices for the random forest algorithms are shown in Tables~\ref{tab:py_bin_cm} and \ref{tab:r_bin_cm}.
In contrast with the multiclass approach, the \textbf{R} and sklearn algorithms return identical confusion matrices.
This is most likely due to the binary case being less noisy and having less complex decision trees.
These confusion matrices also illustrate the gains in accuracy made by grouping all non-XRB categories together - the number of true and false positive XRB sources is similar to the multiclass case.
The overall number of incorrect classifications is reduced compared with the multiclass case where the majority of misclassifications were amongst the three non-XRB classes, which are now correctly classified as simply non-XRB sources. 

\begin{table*}
\caption{Algorithm Evaluation, binary case \label{tab:r_binary_perf}}
\begin{tabular}{lccccc}
\hline\hline
Algorithm & Accuracy & Precision & Recall & AUC$^*$ & CV Score \\
\hline
Logistic Regression     & 0.71 & 0.55 & 0.77 & 0.74 & $0.66\pm0.06$ \\
Naive Bayes             & 0.73 & 0.58 & 0.77 & 0.85 & $0.74\pm0.09$ \\
Support Vector Class.   & 0.71 & 0.55 & 0.77 & 0.85 & $0.71\pm0.08$ \\
Random Forest (sklearn) & 0.84 & 0.71 & 0.85 & 0.88 & $0.75\pm0.05$ \\
Multi-layer Perceptron  & 0.61 & 0.46 & 0.63 & 0.62 & $0.72\pm0.07$ \\
Random Forest (R)       & 0.86 & 0.75 & 0.86 & 0.89 & $0.79\pm0.06$ \\
\hline
\end{tabular}

$^*$Area Under Curve
\end{table*}

\begin{table}
\caption{Confusion matrix for sklearn random forest, binary case \label{tab:py_bin_cm}}
\begin{tabular}{l|l|c|c|c}
\multicolumn{2}{c|}{}&\multicolumn{2}{c}{Actual Class}&\\
\cline{3-4}
\multicolumn{2}{c|}{}&XRB&non-XRB&\multicolumn{1}{c}{Total}\\
\cline{2-4}
\multirow{2}{*}{Predicted Class}& XRB & $15$ & $5$ & $20$\\
\cline{2-4}
& non-XRB & $2$ & $27$ & $29$\\
\cline{2-4}
\multicolumn{1}{c}{} & \multicolumn{1}{c}{Total} & \multicolumn{1}{c}{$17$} & \multicolumn{    1}{c}{$32$} & \multicolumn{1}{c}{$49$}\\
\end{tabular}
\end{table}

\begin{table}
\caption{Confusion matrix for R random forest, binary case \label{tab:r_bin_cm}}
\begin{tabular}{l|l|c|c|c}
\multicolumn{2}{c}{}&\multicolumn{2}{c}{Actual Class}&\\
\cline{3-4}
\multicolumn{2}{c|}{}&XRB&non-XRB&\multicolumn{1}{c}{Total}\\
\cline{2-4}
\multirow{2}{*}{Predicted Class}& XRB & $15$ & $5$ & $20$\\
\cline{2-4}
& non-XRB & $2$ & $27$ & $29$\\
\cline{2-4}
\multicolumn{1}{c}{} & \multicolumn{1}{c}{Total} & \multicolumn{1}{c}{$17$} & \multicolumn{1}{c}{$32$} & \multicolumn{1}{c}{$49$}\\
\end{tabular}
\end{table}

\section{Classification validation by crossmatching}
\label{sec:phat}

To provide an independent method of evaluating our algorithms' classification strength that is not based on X-ray properties, 
we make use of the PHAT \textit{Hubble} imaging survey and the cross-matching between \Chandra\ and \textit{Hubble} 
performed by \citet{Williams18a} and \citet{lazzarini2018}.
\changed{We compare the random-forest classification of previously-unclassified \cite{Vulic16a} sources (i.e., our test set) with
the optical counterpart classifications by defining}
a compatibility score which takes into account the differences between the X-ray and optical classification schemes.
While `foreground star' and `supernova remnant' are classifications in both X-ray and optical schemes, the optical scheme
also includes classes such as `star cluster,' `point source' and `non-detection' that are not used in the X-ray scheme.
There are also categories which are not the same but do overlap: 
if an X-ray source is co-located with a PHAT-identified star cluster or background galaxy, it is overwhelmingly likely that that X-ray source is an X-ray binary or AGN, respectively. 
We consider classifications compatible or incompatible as follows:
\begin{itemize}
    \item X-ray XRBs are compatible with optical point sources, non-detections, star clusters, and objects of unknown type  and incompatible with Hubble galaxies, foreground stars and supernova remnants.
    \item X-ray non-XRBs (in the binary classification) are compatible with all types of Hubble sources except star clusters.
    \item X-ray AGN are compatible with optical point sources, non-detections, galaxies and objects of unknown type, and incompatible with Hubble star clusters, foreground stars and supernova remnants.
    \item X-ray foreground stars are compatible with optical foreground stars and incompatible with all other types.
    \item X-ray supernova remnants are compatible with optical supernova remnants and incompatible with all other types.
\end{itemize}
The compatibility score for each match is computed by summing the number of objects for which the classifications are compatible
and dividing by the number of objects.
\changed{This comparison does not consider the X-ray sources which do not have an optical counterpart; these could be either XRBs or AGN (see \autoref{sec:phat_data}), our two most populated classes, and do not add much discrimination to the comparison.} 

\changed{
We matched the 780 newly classified \cite{Vulic16a} sources with those from \citet{Johnson15a}, \citet{Williams18a} and \citet{lazzarini2018}, using a tolerance of 1.0~arcsec.% 
\footnote{The value of 1.0~arcsec was chosen to reflect the \Chandra\  centroiding precision and astrometric accuracy; see \citet{Williams18a}. 
Changing the exact tolerance had minimal effects on the matching results.}  
We find a total of 10, 14, and 22 matches, respectively; there are five objects in common between the 14 matches with the catalogue of \citet{Williams18a} and the 22 matches with the catalogue of \citet{lazzarini2018}.}
There are no overlaps between the matches with \citet{Johnson15a} and the other two catalogues.
The detailed match results, with object names and coordinates, are given in Appendix~\ref{sec:xmatch_details}. 
For the 10 objects found in the \citet{Johnson15a} cross-match, the compatibility score is 0.9 for the binary random forest classification and 0.8 for the multi-class random forest classification. 
The incompatible objects are 004248.83+411512.9, classified as a foreground star by our method and `unknown' by \citet{Johnson15a} 
and 004325.64+411537.4, classified as non-XRB/AGN by our method and `cluster' by \citet{Johnson15a}.%
\footnote{The surface density of \citet{Johnson15a} clusters is roughly 1.5~arcmin$^{-2}$, so the probability of a chance superposition of cluster and the matching area around an X-ray source is low, about 0.1\%.}
The compatible objects are classified as either XRBs by our method and clusters by  \citet{Johnson15a}, or AGN by our method and `unknown' by \citet{Johnson15a}.
For the 14 objects found in the \citet{Williams18a} cross-match, the compatibility score is 0.86 for both the binary random forest classification and the multi-class random forest classification.
The two incompatible objects are classified (with probabilities 0.51 and 0.57) as XRBs by our method and as galaxies by \citet{Williams18a}
All 22 of the objects found in the \citet{lazzarini2018} cross-match have compatible classifications, so the compatibility score is 1.0.
Removing the double-counting of objects in common between the \citet{Williams18a} and \citet{lazzarini2018} catalogues,
the overall compatibility score is $(8+12+17/10+14+17) = 91$\%.
This is in line with expectations from the binary results, as discussed above in \autoref{sec:binary-results}, though we caution that the number of matched sources is still very small.

\section{Discussion}
\label{sec:discussion}

\begin{figure}
    \centering
    \includegraphics[width=\columnwidth]{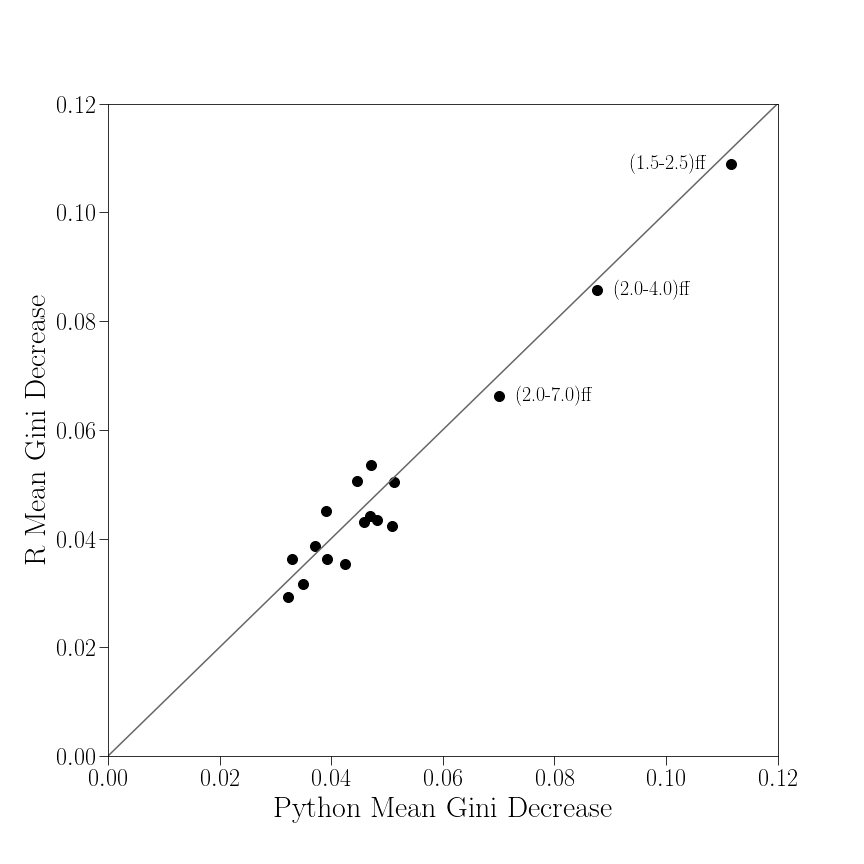}
    \caption{Comparison of feature importances between the \texttt{sklearn} and \textbf{R} implementations of the random forest when the algorithm is trained on the binary realization of the data.
    Labelled datapoints with "ff" indicate that a feature is a photon flux fraction (relative to the total photon flux across the \Chandra\ band) measured in a particular energy range (in keV). 
    Feature importances are ranked using the mean decrease in the Gini coefficient; features with a lower mean decrease are less important to the classifier.}
    \label{fig:binary_importance}
\end{figure}

\begin{figure}
    \centering
    \includegraphics[width=\columnwidth]{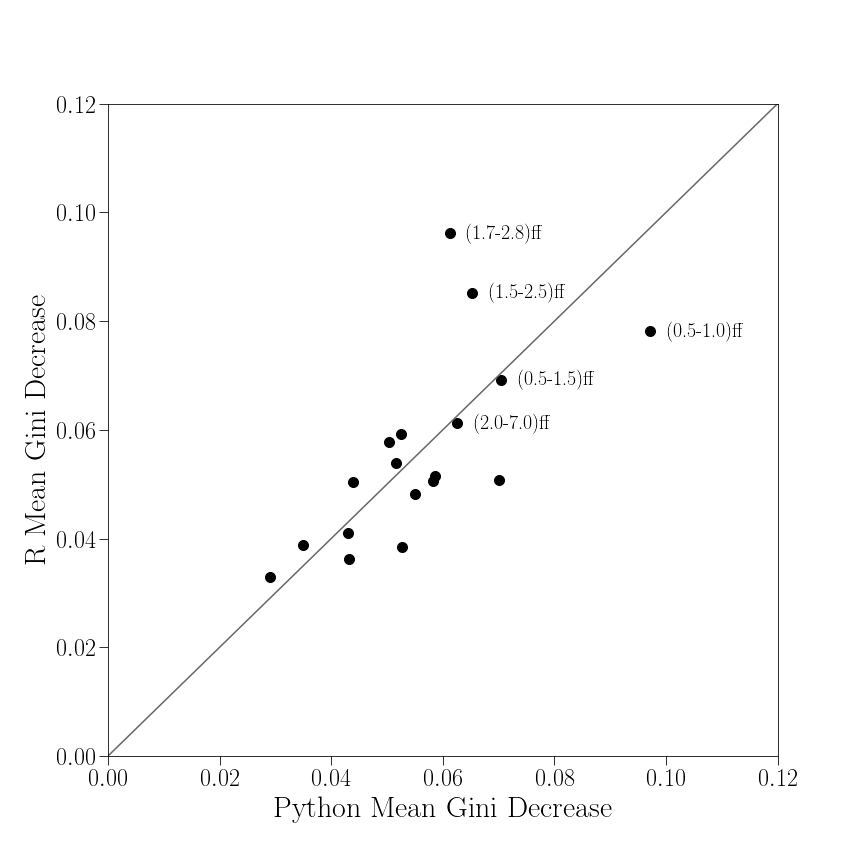}
    \caption{Comparison of feature importances between the \texttt{sklearn} and \textbf{R} implementations of the random forest when the algorithm is trained on the multiclass realization of the data. 
    Labelled datapoints with "ff" indicate that a feature is a photon flux fraction (relative to the total photon flux across the \Chandra\ band) measured in a particular energy range (in keV). 
    Feature importances are ranked using the mean decrease in the Gini coefficient; 
    features with a lower mean decrease are less important to the classifier.} 
    \label{fig:multiclass_importance}
\end{figure}

\begin{figure}
    \centering
   \includegraphics[width=\columnwidth]{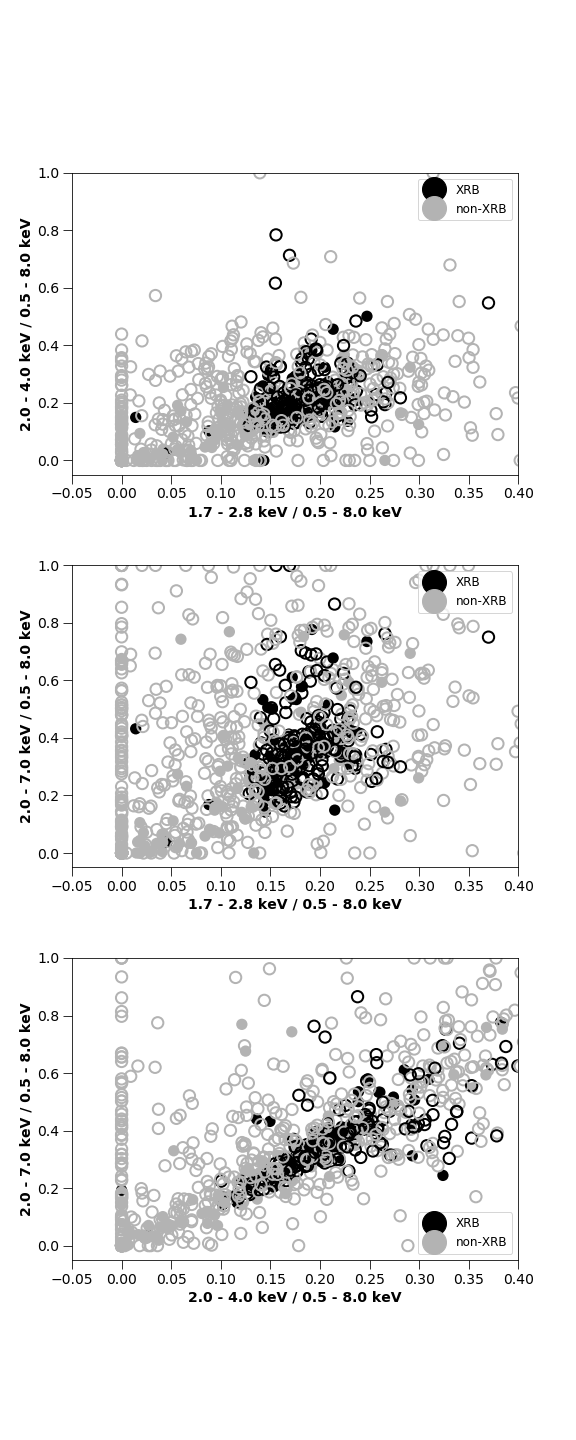}
    \caption{Feature space for three of the most significant features in the binary approach as determined by the \texttt{sklearn} random forest. 
     Filled symbols represent previously classified sources and empty symbols are the sources classified by the algorithm.
    We chose to plot the features that are the most distinct yet still have high significance. The piling up of features at 0.0 and 1.0 on each plot is due to the tacking of values outside of this range to these boundaries.}
    \label{fig:paramspace_binary}
\end{figure}

\begin{figure}
    \centering
    \includegraphics[width=\columnwidth]{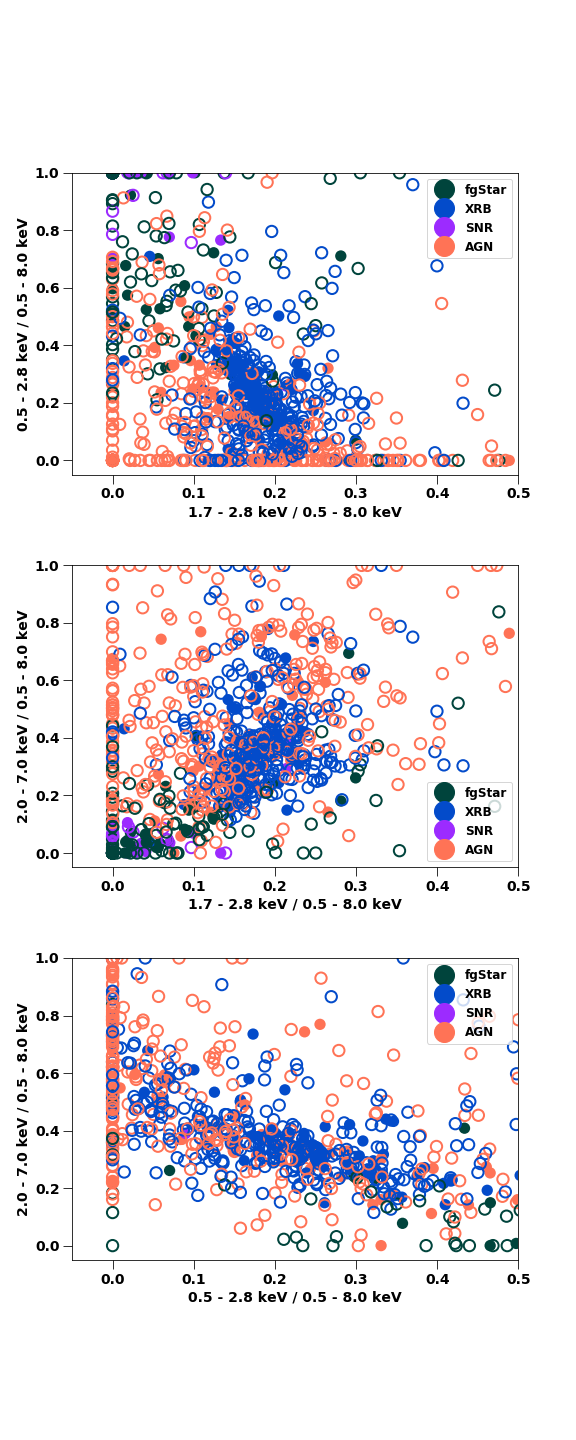}
    \caption{Feature space for three of the most significant features in the multiclass approach as determined by the \texttt{sklearn} random forest.
    Filled symbols represent previously classified sources and empty symbols are the sources classified by the algorithm.
    Since a number of the features determined to be most significant are similar, we chose to plot the features that are the most distinct yet still have high significance. 
    The piling up of features at 0.0 and 1.0 on each plot is due to the tacking of values outside of this range to these boundaries in order for the photon flux fractions to be physically interpretable.  
    }
    \label{fig:paramspace_multiclass}
\end{figure}

We find that algorithms perform significantly better using a binary approach (XRB vs non-XRB) rather than a multiclass approach. 
This is expected for several reasons.
Firstly, we have fairly significant class imbalance and a low number of classified samples overall. 
XRBs are the most numerous class and only a handful of supernova remnants and foreground stars are present by comparison. 
In general, the performance of an algorithm depends strongly on the number of available classified examples; the precise number necessary depends on the structure of the feature space and the desired significance levels \citep{Raudys91a}.
Secondly, classification using only X-ray emission in the narrow energy range of telescopes like \Chandra\ (e.g., 0.5--8.0 keV) is expected to be insufficient as a discriminating classifier based on theoretical models of emission for different X-ray emitters. 
However, our results seem to suggest that for the particular case of separating out XRBs, we may be able to use X-ray information to find the best candidate XRBs.
Traditional methods using X-ray colour-colour diagrams or colour-colour intensity diagrams have hinted at partial separability between XRBs and other kinds of objects, though often with significant overlap and a dependence on the energy range available \citep{Prestwich03a}.
The hard X-ray range of telescopes like \textit{NuSTAR} has been shown to improve separability \citep{Vulic18a}.
Our results suggest that using a higher-dimensional approach with this method may yield more useful results.
In addition, the use of ML techniques does not require the same kind of modelling assumptions as ordinary regression in determining, for example, a linear decision boundary in a colour-colour plot. 
\changed{
Finally, given that our scientific interest is in discriminating XRB candidates from other object types, improving the discrimination between non-XRB types is less important and so a binary classification is sufficient.
}

Across all of our modes of analysis and performance evaluation (accuracy, \changed{precision,} recall, CV score, AUC, and confusion matrices), the random forest algorithms (both sklearn and R) tend to give overall the best performance.
The superior performance of the random forest is unsurprising, given the following properties of our dataset:
\begin{itemize}
    \item Few classified examples ($<200$ total, some categories with fewer than 20 members)
    \item Relatively few features ($<20$, a number of which are linearly dependent on each other)
    \item Features that have differing normalizations (most are ratios but there are 3 features that are bounded differently)
    \item Complex feature space that is unlikely to be linearly separable by category (significant overlap in the feature space between different kinds of object)
\end{itemize}
Logistic regression tends to work well when feature values make independent, additive contributions to class probabilities.
Naive Bayes makes the assumption that features are conditionally independent from each other.
Multi-layer perceptrons often require significant tuning of hyperparameters in order to be useful \citep{sklearnpaper}.
Support vector classifiers do not have these same limitations, however they do not provide direct computation of class membership probabilities (in \texttt{sklearn} these are computed from five-fold cross validation instead). 
The performance of our SVC could likely be improved through tuning of algorithm hyperparameters, including those which account for class imbalance.
\changed{These results are generally consistent with the properties of random forest, support vector, and neural network classifiers as described by \citet{Baron19}.
}

\begin{figure*}
    \centering
    \includegraphics[width=7.5cm]{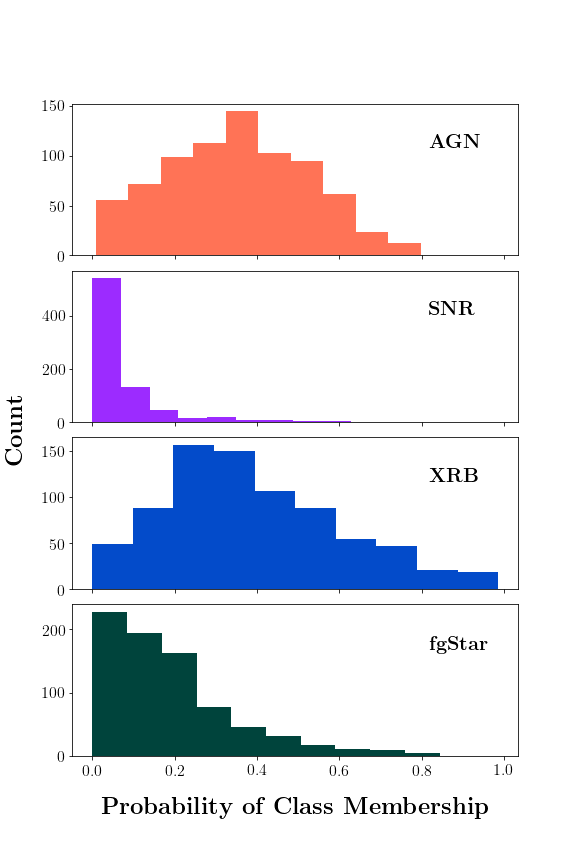}
    \includegraphics[width=7.5cm]{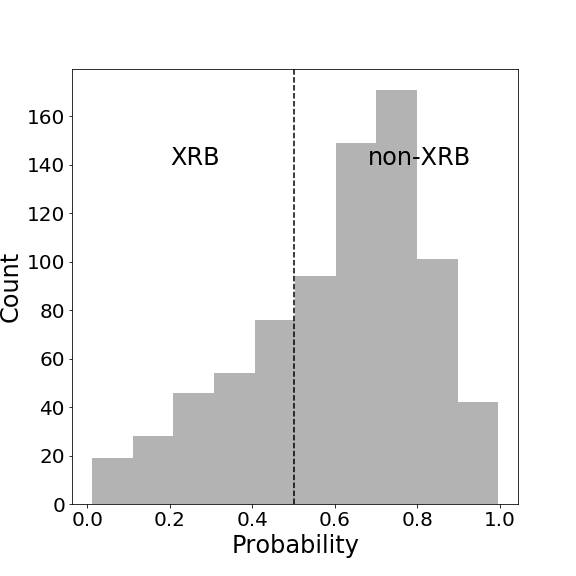}
    \caption
    {Distribution of probability values from random forest classification. 
    Left: multi-class approach.
    Right: binary approach.}
    \label{fig:prob_dist}
\end{figure*}

Using our best performing algorithm, namely the \texttt{sklearn} random forest, the prediction of the 780 unclassified sources using the multiclass approach results in 345 candidate XRBs, 321 candidate background AGNs, 101 candidate foreground stars, and 13 candidate supernova remnants. 
The binary approach identifies only 217 candidate XRBs, while the remaining 563 sources are classified as non-XRB. 
Not all of the candidate XRBs are equally likely - the probability inferred from random forest is based on the fraction of decision trees that vote for a particular classification. 
\autoref{fig:prob_dist} shows the distribution of probability values for the two approaches: 
for each individual object, the probability values sum to 1.0.
In the binary approach, the probability distribution is asymmetric with mean $P({\rm XRB})=0.39$ and standard deviation 0.22,
indicating that the majority of sources are not classified as X-ray binaries.
In the multi-class approach, the probability distributions for the foreground star and supernova remnant classifications peak at low values, indicating that these classes can be ruled out for many objects.
The probability distributions for the AGN and XRB classifications are much broader with means of 0.35 and 0.38, respectively, indicating that distinguishing between these two classes is more difficult. 
$P({\rm XRB})$ is in very good agreement between the binary and multiclass approaches, with a Pearson correlation coefficient of 0.96.

Of the four classes, only the XRB class has probability values above 80\%, which is promising for the goal of this work 
in identifying new X-ray binary candidates. 
Amongst the candidate XRBs identified by the multiclass random forest approach, 19 have a probability of 90\% or greater. 
Of the candidate XRBs in the binary approach, 16 have a probability of XRB classification (as decided by the random forest) of 90\% or greater, all of which are included in the set of 19 high-probability candidates from the multiclass approach.
To facilitate follow-up, these high-probability XRB candidates are listed in \autoref{tab:xrb_cand}.
Four of our high-probability candidates are matched with objects detected by \cite{lazzarini2018} and are discussed further in that work;
the cross-matching information is given in \autoref{tab:phatmatches}.
The difference in the number of high-probability XRB candidates can mostly be explained by the difference in classification thresholds. 
By default, in the binary approach, a source requires 50\% + 1 trees to classify it as an XRB to meet the threshold for XRB classification.
By contrast, in the multiclass approach, a source requires only a plurality of trees to classify as an XRB, which can be a smaller fraction of trees overall. 
The difference in the number of high-probability classifications for the binary versus multiclass approach is due to the binary approach giving more accurate classifications overall. 

\begin{table*}
\caption{High-probability X-ray binary candidates}
\label{tab:xrb_cand}
\begin{tabular}{clllll}
\hline
\hline
ID &  RA &  Dec &  0.5--8 keV photon flux & 0.5--2~keV ff$^*$ &  2--8~keV ff \\
& deg & deg & photons~cm$^{-2}$~s$^{-1}$ & & \\
\hline
004207.73+411814.9 &  10.532222 &   41.304166 &        0.000158 &        0.443604 &        0.437213 \\ 
004210.27+411509.8 &  10.542833 &   41.252749 &        0.000015 &        0.464850 &        0.403821 \\ 
004215.67+411720.7 &  10.565303 &   41.289109 &        0.000037 &        0.466194 &        0.392107 \\ 
004221.48+411601.2 &  10.589510 &   41.267000 &        0.000125 &        0.457008 &        0.410038 \\ 
004222.94+411535.2 &  10.595600 &   41.259787 &        0.000437 &        0.431956 &        0.447427 \\ 
004228.28+411223.1 &  10.617865 &   41.206418 &        0.000204 &        0.451653 &        0.403291 \\ 
004231.13+411621.5 &  10.629732 &   41.272662 &        0.000236 &        0.445122 &        0.401302 \\ 
004232.74+411310.8 &  10.636457 &   41.219675 &        0.000025 &        0.441122 &        0.403111 \\ 
004235.20+412005.6 &  10.646681 &   41.334889 &        0.000046 &        0.420714 &        0.451917 \\ 
004240.20+411845.0 &  10.667528 &   41.312520 &        0.000047 &        0.431413 &        0.428112 \\ 
004242.47+411553.6 &  10.676966 &   41.264899 &        0.000074 &        0.418052 &        0.429165 \\ 
004243.85+411603.8 &  10.682744 &   41.267730 &        0.000028 &        0.459551 &        0.362108 \\ 
004245.11+411621.6 &  10.687963 &   41.272685 &        0.000083 &        0.465262 &        0.352019 \\ 
004246.15+411543.1 &  10.692298 &   41.261986 &        0.000010 &        0.461612 &        0.351838 \\ 
004247.85+411622.9 &  10.699413 &   41.273032 &        0.000006 &        0.428846 &        0.405628 \\ 
004248.52+411521.1 &  10.702178 &   41.255877 &        0.000295 &        0.417619 &        0.426424 \\ 
004249.22+411815.8 &  10.705093 &   41.304405 &        0.000040 &        0.462437 &        0.360194 \\ 
004255.36+412557.4 &  10.730690 &   41.432632 &        0.000095 &        0.432410 &        0.462927 \\ 
004257.90+411104.6 &  10.741271 &   41.184626 &        0.000175 &        0.465306 &        0.361041 \\
\hline
$^*$photon flux fraction
\end{tabular}
\end{table*}

One advantage of the random forest method is that it is straightforward to obtain the relative importance of different features to the final, trained classification algorithm using methods such as the mean decrease in Gini coefficient.
We have plotted this value (normalized to sum to 1 for all features) for the sklearn and \textbf{R} random forests in both the binary and multiclass approach in Figures~\ref{fig:binary_importance} and \ref{fig:multiclass_importance}.
We find that although there is variation between the two algorithms, both sklearn and \textbf{R} tend to weight the same features as important, and in the binary approach (which we expect to be less noisy overall), we find even stronger agreement.

The parameter spaces for a few of the most important features in the multiclass and binary cases are shown in Figures~\ref{fig:paramspace_binary} and \ref{fig:paramspace_multiclass}.
In both cases a number of the features determined to be most significant are similar, so we chose to plot the features that are the most distinct yet still have high significance. 
As expected, even when plotting the most significant features, there is no clear separating boundary between the categories of sources. 
In the multiclass plots, supernova remnants and foreground stars tend to have large fractions in the softest bands (those that begin at the 0.5~keV edge of the \textit{Chandra} range), while AGN are found throughout the parameter space.
XRBs and XRB candidates also occur throughout the parameter space, however they tend to cluster at intermediate values of these soft bands. In the binary case, the strip in the parameter spaces occupied by XRBs is narrower than in the multiclass approach. 
The binary approach-classified XRBs and XRB candidates tend to have some flux in the harder bands (e.g., 2.0--7.0~keV), but they tend neither to be the softest nor the hardest sources in the sample. 

Curiously, we find that the features judged to be strongly predictive are flux ratios obtained from bands that are less common in traditional hardness ratio analyses, and are not generally measured for \Chandra\ datasets except in the ACIS EXTRACT defaults.
Some of these bands, such as 1.7--2.8 keV or 0.5--1.0 keV, tend to be narrower than typical cuts made for flux ratios, whereas others, such as 2.0--4.0 keV and 2.0--7.0 keV, are atypical cuts made for flux ratios even if they are relatively wide slices of the \Chandra\ energy range.
Narrower bands are expected to be less useful overall since more counts are needed in order to measure the flux in these bands accurately. 
Detailed interpretation of the significance of these bands is deferred to future work, though we briefly place the bands selected in context here.

\cite{Prestwich03a} plotted different categories of X-ray source in nearby galaxies (including M31's bulge) observed with \textit{Chandra} using hardness ratios with separations at 0.3--1.0 keV, 1.0--2.0 keV, and 2.0--10.0 keV. 
Several properties of X-ray sources are noted.
LMXBs tend to have spectra described by a power law of photon index 1.5--2 combined with intrinsic absorption.
HMXBs appear harder in the \textit{Chandra} range, with an index of 1--2, though there is a dependence on the accretor type; neutron star XRBs typically have harder spectra than black hole XRBs due to neutron stars having a solid surface  \citep{Binder2015}.
Both classes can have emission above 10 keV, though LMXBs and HMXBs are expected to peak above and below 15 keV.

Combined, both classes of XRB are intrinsically harder than supernova remnants, which have soft spectra peaking below 2 keV because of a combination of shock-heated plasma and atomic emission lines \citep{Yokogawa03a}.
X-ray active foreground stars, likewise, also have spectra dominated by optically thin thermal plasma and are expected to show up as relatively soft X-ray emitters \citep{Ducci13a}. 
It is likely that low values for the 2.0--7.0 keV and 2.0--4.0 keV bands and high values for the 0.5--1.0 keV or 1.7--2.8 keV bands are being driven by the intrinsically soft SNR or foreground star X-ray spectra.
An interesting future study would be to combine SNRs and foreground stars into a single category and use a three-class approach to training and classification. 
\cite{Binder2015} used this approach on the basis of their similarly expected properties to classify X-ray sources in NGC 55, NGC 2403, and NGC 4214.
Notably, their Bayesian approach also included source position relative to the galaxy, which was not available to us since our source coverage is limited to a handful of fields within M31. 

The comparison of AGN to XRBs presents a more complicated picture.
AGN are described by a power law continuum with index $\Gamma = 1.7-2$ in soft X-rays, with additional contributions from non-thermal inverse Compton scattering of accretion disk photons by hot electrons above the disk, photoelectric absorption edges due to gas along the line of sight, and relativistically broadened emission lines around 6~keV \citep{Nandra94a,George91a,Morrison83a,Fabian2000a}.
It is possible that the relative importance of the 2.0--7.0 keV band in our AGN/XRB discriminiation is due to the relativistically broad iron line present in AGN \changed{X-ray spectra}.

\section{Conclusions and Future Work}

We have constructed a proof-of-concept method for improving classification of X-ray sources in nearby galaxies using machine learning techniques.
Our results are summarized as follows:

\begin{itemize}
    \item After testing a variety of algorithms, we find that random forest classification tends to outperform other classifiers, offering an accuracy of $\sim85$\% at separating X-ray binaries from other kinds of contaminating X-ray sources. \changed{This classification scheme tends to deliver the best precision, which is preferred for computing a population's X-ray luminosity function. It also has the best recall, ideal for identifying new XRB candidates. }
    \item Using a binary approach to classification, we find 16 new strong (probability $>90$\%) XRB candidates that are suitable as candidates for followup.
    \item Cross-matching previously unclassified X-ray sources with sources classified using the optical PHAT survey, we find compatible classifications for 37 of 41 objects.
    \item The X-ray bands which tend to rank highest in importance for classification are typically narrower and/or less commonly used bands, such as the 1.7--2.8, 0.5--1.0, 2.0--4.0, and 2.0--7.0~keV photon flux ratios.
\end{itemize}

A primary limitation of machine learning techniques is that they tend to offer poor predictive performance for small sample sizes.
In our sample, we have fewer than 200 total classified examples, and there is significant imbalance between the four types of X-ray source identified by previous surveys.
The number of sources required for accurate classification depends on the desired significance threshold, amongst other parameters \citep{Beleites12a}. 
Additionally, classified samples to use as input for ML methods in astronomical data are typically those objects which are brightest, nearest, and have the longest duty cycles, which may impact new, unclassified samples, if they are distributed in areas of the parameter space where there are few examples available for classification \citep{Richards06a}.

The most obvious pathway for improvement of our methodology is to include more classified examples in the algorithm training.
More broadly, a more useful version of this algorithm would be able to predict X-ray classification for any X-ray source, regardless of expected population. 
Promisingly, our best performance RF algorithms do not strongly weight the only distance-dependent feature in our feature vector (0.5--8.0~keV photon flux) as a strong determinant in classifying X-ray sources.
As such, combining results from surveys of well-studied nearby galaxies will improve our detection algorithm, though caution must be taken.
For example, the effect of Galactic extinction along different lines of sight may necessitate additional corrections for absorption so that X-ray sources from different galaxies can be compared to each other.
Determining average fluxes from multi-epoch observations may also bias classifications.
Other effects which we could take into account for future trained versions of this algorithm would include treatment of uncertainty on the flux ratios used as features, corrections for variability, and consideration of other features such as spatial location relative to galaxy structure and the properties of optical counterparts (e.g, magnitudes in the various PHAT filters for the M31 dataset) as features. 
When compared with XRBs and AGNs, neither foreground stars or SNRs show rapid variability \citep{Binder2015}, so a feature which characterizes variability over all observations may be an additional discriminant. 
Large catalogues of X-ray sources, such as the \Chandra\ Source Catalog \citep{Evans10a}, may also provide a mineable source of classified examples.

We could also improve our classification strength by performing a more detailed investigation of algorithm performance for different values of the algorithm hyperparameters.
A deeper study would involve conducting a gridded search that varies all algorithm hyperparameters at once and evaluates algorithm performance at each value using cross-validation or a similar performance metric. 
Although each particular algorithm can be fit to a particular training set quickly, the number of fits required to do a grid search scales as $\prod_{i=1}^{n} f(i)$, where $i$ is the number of hyperparameters and $f(i)$ is the number of values tested for the $i^{\rm th}$ hyperparameter.
This can easily become computationally time-consuming to evaluate for all of the algorithms presented in this work.
However, a potential future extension would be to perform a more detailed study of the hyperparameter space of the random forest algorithm, since it offers the best performance in this classification task.
As discussed above in \autoref{sec:multi-results}, the main source of poor classification accuracy in the multi-class approach is misclassifications of AGN combined with low numbers of foreground stars and supernova remnants. 
A next generation of ML algorithm which includes multiwavelength properties would hopefully rectify this issue -- foreground stars, for example, can often be ruled out on the basis of their extreme optical/X-ray flux ratios.
Improved classification and identification of these unknown X-ray sources will enable better understanding of the populations of XRBs inside of galaxies, and may also provide clues about the nature of XRB emission.

\section*{Acknowledgements}

We thank the referee for helpful comments.
R. M. A. acknowledges support from an NSERC CGS-D scholarship. 
P. B. acknowledges support from an NSERC Discovery Grant.
N. V. acknowledges funding from Ontario Graduate Scholarships.
We thank E. Cackett, S. Gallagher, M. Gorski, D. Lizotte, and T.A.A. Sigut for helpful discussions.
The scientific results reported in this article are based in part on observations made by the Chandra X-ray Observatory. 

\bibliographystyle{mnras}
\bibliography{M31_ML.bib}

\appendix

\section{Crossmatching details}
\label{sec:xmatch_details}

The results of cross-matching the unidentified X-ray sources from \cite{Vulic16a} with the catalogues from the Andromeda Project \citep{Johnson15a}
are given in \autoref{tab:andprj_matches}.
Cross-matching results with the \Chandra-PHAT survey \citep{Williams18a} and follow-up by \citet{lazzarini2018}
are given in \autoref{tab:phatmatches}.
In these tables, $P_B({\rm XRB})$ and $P_M({\rm XRB})$ give the probability that the source is an X-ray binary, based on the
random forest binary and multi-class approches, respectively.
These results are further discussed in the text in \autoref{sec:phat}.

\begin{table*}
\caption{Andromeda Project Crossmatches to Unidentified X-ray Sources}
\label{tab:andprj_matches}
\begin{tabular}{lllllllrl}
\hline
\hline
ID & RA & Dec & RF binary XRB? & $P_B({\rm XRB})$ & RF multiclass & $P_M({\rm XRB})$ & AndProj ID & AndProj type \\
\hline
004233.25+411742.2 & 10.638557 & 41.295061 & yes & 0.842 & XRB & 0.850 & 5002 & cluster \\ 
004246.08+411736.1 & 10.692008 & 41.293385 & yes & 0.784 & XRB & 0.786 & 8052 & cluster \\ 
004248.83+411512.9 & 10.703484 & 41.253587 & no & 0.03 & fgStar & 0.000 & 12930 & unknown \\ 
004250.81+411707.3 & 10.711741 & 41.285387 & yes & 0.566 & XRB & 0.556 & 6947 & cluster \\
004255.60+411835.0 & 10.731686 & 41.309739 & yes & 0.742 & XRB & 0.740 & 7980 & cluster \\ 
004325.64+411537.4 & 10.856839 & 41.260392 & no & 0.19 & AGN & 0.294 & 2927 & cluster \\ 
004356.59+410644.3 & 10.985825 & 41.112333 & no & 0.258 & AGN & 0.352 & 6784 & unknown \\
004614.67+414317.6 & 11.561131 & 41.721574 & no & 0.312 & AGN & 0.320 & 11394 & unknown \\ 
004615.36+414128.1 & 11.564025 & 41.691153 & no & 0.298 & AGN & 0.330 & 9397 & unknown \\ 
004616.82+414300.4 & 11.570103 & 41.716791 & no & 0.098 & AGN & 0.172 & 11964 & unknown \\ 
\hline
\end{tabular}
\end{table*}

\begin{landscape}

\begin{table}
\caption{Chandra/NuSTAR-PHAT Crossmatches to Unidentified X-ray Sources}
\label{tab:phatmatches}
\begin{tabular}{llllllllll}
\hline
\hline
ID & RA & Dec & RF binary XRB? & $P_B({\rm XRB})$ & RF multiclass & $P_M({\rm XRB})$ & ChandraPHAT ID & NuStar ID & PHAT type$^*$ \\
\hline
004220.93+411520.1 & 10.587243 & 41.255601 & no & 0.376 & AGN & 0.352 & & 004220.96+411520.3 & n \\
004235.20+412005.6 & 10.646681 & 41.334889 & yes & 0.884 & XRB & 0.904 &  & 004235.20+412005.0 & n \\
004246.15+411543.1 & 10.692298 & 41.261986 & yes & 0.956 & XRB & 0.946 &  & 004246.19+411543.2 & n \\
004248.52+411521.1 & 10.702178 & 41.255877 & yes & 0.988 & XRB & 0.986 &  & 004248.56+411520.8 & n \\
004249.22+411815.8 & 10.705093 & 41.304405 & yes & 0.882 & XRB & 0.920 &  & 004249.22+411815.5 & p \\
004254.92+411603.1 & 10.728856 & 41.267551 & yes & 0.9 & XRB & 0.886 &  & 004254.93+411602.8 & n \\
004255.17+411836.0 & 10.729879 & 41.310007 & yes & 0.648 & XRB & 0.668 &  & 004255.19+411835.4 & n \\
004255.60+411835.0 & 10.731686 & 41.309739 & yes & 0.742 & XRB & 0.740 &  & 004255.60+411834.5 & c \\
004303.00+412041.7 & 10.762519 & 41.344936 & no & 0.466 & XRB & 0.478 &  & 004303.03+412041.6 & n \\
004311.35+411809.5 & 10.797296 & 41.302656 & no & 0.352 & AGN & 0.348 &  & 004311.37+411809.3 & n \\
004313.84+411711.2 & 10.807687 & 41.286457 & no & 0.438 & XRB & 0.424 &  & 004313.88+411711.5 & n \\
004321.06+411750.5 & 10.837783 & 41.297370 & yes & 0.688 & XRB & 0.668 &  & 004321.07+411750.2 & p \\
004321.52+411557.3 & 10.839692 & 41.265932 & no & 0.312 & AGN & 0.350 &  & 004321.48+411556.5 & p \\
004324.83+411726.6 & 10.853461 & 41.290744 & yes & 0.608 & XRB & 0.514 &  & 004324.84+411726.9 & n \\
004332.37+411040.8 & 10.884898 & 41.178020 & yes & 0.624 & XRB & 0.628 &  & 004332.38+411040.9 & n \\
004334.31+411323.5 & 10.892964 & 41.223215 & yes & 0.712 & XRB & 0.714 &  & 004334.33+411323.1 & n \\
004335.84+411433.6 & 10.899342 & 41.242673 & no & 0.334 & AGN & 0.318 &  & 004335.91+411433.0 & p \\
004350.75+412117.5 & 10.961462 & 41.354878 & yes & 0.622 & XRB & 0.612 & 004350.76+412118.1 & 004350.76+412117.4 & p \\
004359.82+412435.4 & 10.999275 & 41.409840 & no & 0.168 & AGN & 0.244 & 004359.83+412435.6 &  & p \\
004401.04+412808.2 & 11.004338 & 41.468952 & no & 0.47 & XRB & 0.468 & 004401.02+412808.8 &  & n \\
004412.20+413148.2 & 11.050855 & 41.530057 & no & 0.428 & AGN & 0.422 & 004412.17+413148.4 &  & p \\
004425.73+412241.7 & 11.107221 & 41.378275 & yes & 0.614 & XRB & 0.550 & 004425.73+412242.4 & 004425.73+412241.8 & p \\
004429.70+412257.5 & 11.123765 & 41.382641 & no & 0.224 & AGN & 0.206 & 004429.73+412258.0 & 004429.73+412257.4 & n \\
004430.47+412310.2 & 11.126980 & 41.386190 & no & 0.466 & AGN & 0.454 & 004430.46+412310.7 & 004430.45+412310.1 & n \\
004448.15+412247.1 & 11.200641 & 41.379767 & yes & 0.576 & XRB & 0.544 & 004448.13+412247.9 & 004448.13+412247.4 & p \\
004542.90+414312.6 & 11.428779 & 41.720189 & yes & 0.546 & XRB & 0.580 & 004542.91+414312.9 &  & n \\
004552.93+414441.8 & 11.470551 & 41.744965 & no & 0.452 & XRB & 0.458 & 004552.98+414441.7 &  & n \\
004555.72+414551.8 & 11.482172 & 41.764389 & yes & 0.574 & XRB & 0.500 & 004555.67+414551.9 &  & g \\
004559.07+414113.0 & 11.496132 & 41.686945 & yes & 0.516 & XRB & 0.520 & 004559.03+414112.7 &  & g \\
004602.70+413856.7 & 11.511251 & 41.649095 & no & 0.232 & AGN & 0.278 & 004602.74+413856.5 &  & n \\
004611.46+413940.1 & 11.547781 & 41.661161 & no & 0.276 & AGN & 0.252 & 004611.46+413940.5 &  & n \\
\hline
\end{tabular}

$^*$c: star cluster, g: galaxy, n: no optical counterpart, p: point source.
\end{table}
\end{landscape}

\bsp	% typesetting comment
\label{lastpage}
\end{document}